\documentclass[twocolumn,aps,superscriptaddress,showpacs]{revtex4}
\usepackage{amsmath,bm}
\usepackage{graphicx}

\begin{document}

\title{Differential isospin-fractionation in dilute asymmetric nuclear matter%
}
\author{Bao-An Li}
\affiliation{Department of Physics, Texas A\&M University-Commerce, Commerce, TX
75429-3011, USA}
\author{Lie-Wen Chen}
\affiliation{Department of Physics, Texas A\&M University-Commerce, Commerce, TX
75429-3011, USA}
\affiliation{Institute of Theoretical Physics, Shanghai Jiao Tong University, Shanghai
200240, China}
\affiliation{Center of Theoretical Nuclear Physics, National Laboratory of Heavy-Ion
Accelerator, Lanzhou, 730000, China}
\author{Hong-Ru Ma}
\affiliation{Institute of Theoretical Physics, Shanghai Jiao Tong University, Shanghai
200240, China}
\author{Jun Xu}
\affiliation{Institute of Theoretical Physics, Shanghai Jiao Tong University, Shanghai
200240, China}
\author{Gao-Chan Yong}
\affiliation{Department of Physics, Texas A\&M University-Commerce, Commerce, TX
75429-3011, USA}
\affiliation{Institute of Modern Physics, Chinese Academy of Sciences, Lanzhou 730000, China}

\begin{abstract}
The differential isospin-fractionation (IsoF) during the liquid-gas
phase transition in dilute asymmetric nuclear matter is studied as a
function of nucleon momentum. Within a self-consistent thermal model
it is shown that the neutron/proton ratio of the gas phase becomes
{\it smaller} than that of the liquid phase for energetic nucleons,
although the gas phase is overall more neutron-rich. Clear
indications of the differential IsoF consistent with the thermal
model predictions are demonstrated within a transport model for
heavy-ion reactions. Future comparisons with experimental data will
allow us to extract critical information about the momentum
dependence of the isovector strong interaction.

\end{abstract}

\maketitle

Properties of isospin asymmetric nuclear matter play an important role in
understanding many key issues in both nuclear physics and astrophysics \cite%
{ireview98,da02,Lat04,steiner05a}. To explore these properties has
been one of the major objectives of nuclear sciences. Recent
progress and new plans in conducting experiments with more advanced
radioactive beams provided us a great opportunity to achieve this
goal. Meanwhile, impressive advances have also been made recently in
several theoretical fronts in exploring the nature of neutron-rich
nucleonic matter and their experimental manifestations. One of the
especially interesting new features of a dilute asymmetric nuclear
matter is the isospin-fractionation (IsoF) during the liquid-gas
(LG) phase transition in it~\cite{muller95,liko97,baran98,li00}. The
non-equal partition of the system's isospin asymmetry with the gas
phase being more neutron-rich than the liquid phase has been found
as a general phenomenon using essentially all thermodynamical models
and in simulations of heavy-ion reactions, for reviews see, e.g.,
Refs. \cite{ireview98,ibook,chomaz,das,Bar05,wci}.

In fact, indications of the IsoF have been reported ever since the early
80's although their interpretations have not always been unique \cite%
{randrup,chomaz}. Several recent experiments confirmed unambiguously
the IsoF phenomenon \cite{wci}. In particular, the experiments and
analyses by Xu et al. \cite{xu} at the NSCL/MSU are among the most
interesting and detailed ones available so far. In their experiments
the isotope, isotone and isobar ratios were utilized to obtain an
estimate of the neutron/proton density ratio $\rho _{n}/\rho _{p}$
in the gas phase at the breakup stage of the reaction. It was found
clearly that the gas phase was significantly enriched in neutrons
relative to the liquid phase represented by bound nuclei. However,
in all of the existing theoretical and experimental studies in the
literature, only the average neutron/proton ratios integrated over
the nucleon momentum in the liquid and gas phases were studied. We
normally refer the above predicted and observed
isospin-fractionation as the integrated IsoF. In this work, we
investigate the differential IsoF as a function of nucleon momentum.
Surprisingly, completely new and very interesting physics is
revealed from the fine structure of the differential IsoF analysis.
\begin{figure}[tbh]
\includegraphics[scale=0.6]{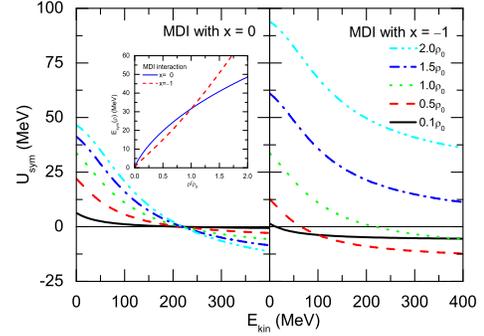}
\caption{{\protect\small (Color online) The symmetry potential and energy
(insert) in the MDI interaction with $x=0$ and $x=-1$.}}
\label{fig1}
\end{figure}

Our study is carried out first within a self-consistent thermal model~\cite%
{xu07} using the isospin and momentum-dependent MDI interaction~\cite%
{das03,chen05}. With this interaction, the potential energy density $%
V(\rho ,T,\delta )$ at total density $\rho $, temperature $T$ and
isospin asymmetry $\delta \equiv (\rho _{n}-\rho _{p})/\rho $ is
\begin{eqnarray}\label{MDIV}
&&V(\rho ,T,\delta ) =\frac{A_{u}\rho _{n}\rho _{p}}{\rho _{0}}+\frac{A_{l}}{%
2\rho _{0}}(\rho _{n}^{2}+\rho _{p}^{2})+\frac{B}{\sigma
+1}\frac{\rho ^{\sigma +1}}{\rho _{0}^{\sigma }}  \\ &&\cdot
(1-x\delta ^{2})+\sum_{\tau ,\tau ^{\prime }}\frac{C_{\tau ,\tau
^{\prime }}}{\rho_0}
\int \int d^{3}pd^{3}p^{\prime }\frac{f_{\tau }(\vec{r},\vec{p}%
)f_{\tau ^{\prime }}(\vec{r},\vec{p}^{\prime
})}{1+(\vec{p}-\vec{p}^{\prime })^{2}/\Lambda ^{2}}.\notag
\end{eqnarray}
In the mean field approximation, Eq. (\ref{MDIV}) leads to the following
single particle potential for a nucleon with momentum $\vec{p}$ and isospin $%
\tau $
\begin{eqnarray}
&&U_{\tau }(\rho ,T,\delta ,\vec{p},x)=A_{u}(x)\frac{\rho _{-\tau }}{\rho
_{0}}+A_{l}(x)\frac{\rho _{\tau }}{\rho _{0}}  \notag  \label{mdi} \\
&&+B\left( \frac{\rho }{\rho _{0}}\right) ^{\sigma }(1-x\delta ^{2})-8\tau x%
\frac{B}{\sigma +1}\frac{\rho ^{\sigma -1}}{\rho _{0}^{\sigma }}\delta \rho
_{-\tau }  \notag \\
&&+\sum_{t=\tau ,-\tau }\frac{2C_{\tau ,t}}{\rho _{0}}\int d^{3}\vec{p}%
^{\prime }\frac{f_{t}(\vec{r},\vec{p}^{\prime })}{1+(\vec{p}-\vec{p}^{\prime
})^{2}/\Lambda ^{2}}.  \label{gogny}
\end{eqnarray}%
where $\tau =1/2$ ($-1/2$) for neutrons (protons), $x$, $A_{u}(x)$,
$A_{\ell }(x)$, $B$, $C_{\tau ,\tau }$,$C_{\tau ,-\tau }$, $\sigma
$, and $\Lambda $ are all parameters given in Ref. \cite{das03}.
This interaction gives an incompressibility of $K_{0}=211$ MeV for
symmetric nuclear matter at saturation density $\rho _{0}=0.16$
fm$^{-3}$. The different $x$ values in the MDI interaction are
introduced to vary the density dependence of the nuclear symmetry
energy while keeping other properties of the nuclear equation of
state fixed \cite{das03,chen05}. It is worth mentioning that the
nucleon isoscalar potential estimated from
$U_{\text{isoscalar}}\approx (U_{n}+U_{p})/2$ agrees with the
prediction of variational many-body calculations for symmetric
nuclear matter \cite{wiringa} in a broad density and momentum range
\cite{das03}. The corresponding isovector (symmetry) potential can be estimated from $%
U_{sym}\approx (U_{\text{n}}-U_{\text{p}})/2\delta $. With both $x=0$ and $%
x=-1$ at normal nuclear matter density, the symmetry potential agrees very
well with the Lane potential extracted from nucleon-nucleus and (n,p) charge
exchange reactions available for nucleon kinetic energies up to about $100$
MeV \cite{das03,data}. At abnormal densities and higher nucleon energies,
however, there is no experimental constrain available at present.

The isospin diffusion data from the NSCL/MSU \cite{betty04} have recently
constrained the value of $x$ to be between $0$ and $-1$ at sub-saturation
densities \cite{chen05,lichen05}. Shown in the insert of Fig. \ref{fig1} is
the density dependent symmetry energy $E_{\mathrm{sym}}(\rho )$ with $x=0$
and $x=-1$, respectively. While this constraint on the $E_{\mathrm{sym}%
}(\rho )$ is the most stringent achieved so far in the field, the
corresponding symmetry potential shown in Fig.\ \ref{fig1} still diverges
much more widely with both momentum and density. This is not surprising.
Comparing Eqs. (\ref{MDIV}) and (\ref{gogny}), it is seen that the symmetry
energy obtained using Eq.\ (\ref{MDIV}) involves the integration of the
single nucleon potential over its momentum. To obtain information about the
underlying momentum- and density-dependence of the symmetry potential which
is more fundamental than the $E_{\mathrm{sym}}(\rho )$ for many important
physics questions, one thus has to use differential probes. We will
demonstrate that the differential IsoF as a function of nucleon momentum is
such an observable.

The phase space distribution function $f_{\tau }$ is the Fermi distribution
\begin{equation}
f_{\tau }(\vec{r},\vec{p})=\frac{2}{h^{3}}[\exp (\frac{\frac{p^{2}}{%
2m_{_{\tau }}}+U_{\tau }-\mu _{\tau }}{T})+1]^{-1}  \label{f}
\end{equation}%
where $\mu _{\tau }$ is the proton or neutron chemical potential determined
self-consistently from $\rho _{\tau }=\int f_{\tau }(\vec{r},\vec{p})d^{3}p.$
In the above, $m_{_{\tau }}$ is the proton or neutron mass. From a
self-consistency iteration scheme \cite{xu07,gale90}, the chemical potential
$\mu _{\tau }$ and the distribution function $f_{\tau }(\vec{r},\vec{p})$
are determined numerically.

From the chemical potential $\mu _{\tau }$ and the distribution function $%
f_{\tau }(\vec{r},\vec{p})$, the energy per nucleon $E(\rho ,T,\delta )$ can
be obtained as
\begin{equation}
E(\rho ,T,\delta )=\frac{1}{\rho }\left[ V(\rho ,T,\delta )+{\sum_{\tau }}%
\int d^{3}p\frac{p^{2}}{2m_{\tau }}f_{\tau }(\vec{r},\vec{p})\right] .
\label{E}
\end{equation}%
Furthermore, the entropy per nucleon $S_{\tau }(\rho ,T,\delta )$ is
obtained from
\begin{equation}
S_{\tau }(\rho ,T,\delta )=-\frac{8\pi }{{\rho }h^{3}}\int_{0}^{\infty
}p^{2}[n_{\tau }\ln n_{\tau }+(1-n_{\tau })\ln (1-n_{\tau })]dp  \label{S}
\end{equation}%
with the occupation probability $n_{\tau }=\frac{h^{3}}{2}f_{\tau }(\vec{r},%
\vec{p})$. Finally, the pressure $P(\rho ,T,\delta )$ is calculated from the
thermodynamic relation
\begin{equation}
P(\rho ,T,\delta )=\left[ T{\sum_{\tau }}S_{\tau }(\rho ,T,\delta )-E(\rho
,T,\delta )\right] \rho +\sum_{\tau }\mu _{\tau }\rho _{\tau }.  \label{P}
\end{equation}

For the ease of following discussions, we first reproduce the integrated
IsoF using the above formalism. In hot asymmetric nuclear matter, the LG
phase coexistence is governed by the Gibbs conditions
\begin{eqnarray}
\mu _{i}^{L}(T,\rho _{i}^{L}) &=&\mu _{i}^{G}(T,\rho _{i}^{G}),(i=n\text{
and }p)  \label{coexistencemu} \\
P_{i}^{L}(T,\rho _{i}^{L}) &=&P_{i}^{G}(T,\rho _{i}^{G}),(i=n\text{ or }p).
\label{coexistenceP}
\end{eqnarray}%
At any given pressure below the critical point, the two solutions for the
liquid and gas phases thus form the edges of a rectangle in the proton and
neutron chemical potential isobars as a function of isospin asymmetry $%
\delta $ and can be found by means of the standard geometrical construction
method \cite{xu07}.
\begin{figure}[tbh]
\includegraphics[scale=0.6]{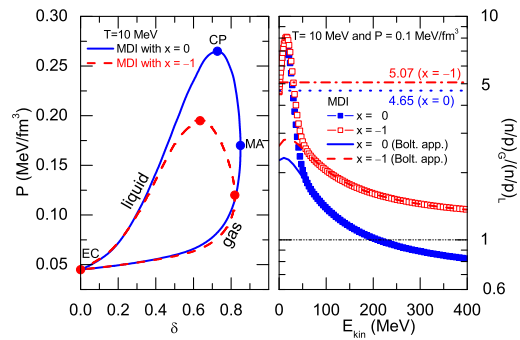}
\caption{{\protect\small (Color online) Left window: the section of binodal surface at $%
T=10$ MeV with $x=0$ and $x=-1$. The critical point (CP), the points of
equal concentration (EC) and maximal asymmetry (MA) are also indicated.
Right window: the double neutron/proton ratio in
the gas and liquid phases $(n/p)_{G}/(n/p)_{L}$ as a function of the nucleon
kinetic energy.}}
\label{fig2}
\end{figure}

Shown in the left window of Fig.~\ref{fig2} is a typical section of the binodal surface at $%
T=10 $ MeV with $x=0$ and $x=-1$. The phenomenon of integrated IsoF with the
gas phase being more neutron-rich is seen very clearly. It is also seen that
the stiffer symmetry energy ($x=-1$) significantly lowers the critical point
(CP). However, below a pressure of about $P=0.12\text{ MeV}$/fm$^{3}$, the
magnitude of the integrated IsoF becomes almost independent of the symmetry
energy used. To examine the advantages of the differential IsoF analyses
over the integrated ones, as an example, we select the gas and liquid phases
in equilibrium at T=10 MeV and $P=0.1\text{ MeV}$/fm$^{3}$. For $x=0$ the
density and isospin asymmetry are, respectively, $\rho _{G}=0.087\rho _{0}$
and $\delta _{G}=0.791$ for the gas phase, and $\rho _{L}=0.763\rho _{0}$
and $\delta _{L}=0.296$ for the liquid phase. For $x=-1$ they are,
respectively, $\rho _{G}=0.114\rho _{0}$, $\delta _{G}=0.808$, $\rho
_{L}=0.714\rho _{0}$ and $\delta _{L}=0.30$. The corresponding double
neutron/proton ratio in the gas and liquid phases $\frac{(n/p)_{G}}{(n/p)_{L}%
}(p)$ can then be readily obtained using Eq. (\ref{f}) as a function of
nucleon momentum or kinetic energy. We refer this function as the
differential IsoF.

Shown in the right window of Fig.~\ref{fig2} are the differential
IsoFs for both $x=0$ and $x=-1$. 
It is interesting to see that the isospin-fractionation is
strongly momentum dependent. Moreover, while the integrated double
neutron/proton ratios of $5.07$ ($x=-1$) and $4.65$ ($x=0$) are
very close to each other, the differential IsoF for nucleons with
kinetic energies high than about $50$ MeV is very sensitive to the
parameter $x$ used. Surprisingly, a reversal of the normal IsoF is
seen for $x=0$ for nucleons with kinetic energies higher than
about $220$ MeV. In this case, there are more energetic neutrons
than protons in the liquid phase compared to the gas phase. We
note that at pressures higher than $0.1$ $\text{MeV}$/fm$^{3}$
where the integrated IsoF is already very sensitive to the $E_{\mathrm{sym}%
}(\rho )$, the differential IsoF is much more sensitive to the $x$
parameter than that shown in Fig.~\ref{fig2}. For energetic
nucleons where the differential IsoF is very sensitive to the
parameter $x$, the $f_{\tau }$ can be well approximated by the
Boltzmann distribution as shown in Fig.~\ref{fig2}. For these
nucleons in either the liquid ($L$) or gas ($G$) phase, the
neutron/proton ratio
\begin{equation}
(n/p)_{L/G}=\exp [-(E_{n}^{L/G}-E_{p}^{L/G}-\mu _{n}^{L/G}+\mu
_{p}^{L/G})/T].
\end{equation}%
The energy difference of neutrons and protons having the same kinetic energy
and mass
\begin{equation}
E_{n}^{L/G}-E_{p}^{L/G}=U_{n}^{L/G}-U_{p}^{L/G}\approx 2\delta _{L/G}\cdot
U_{sym}(p,\rho _{L/G})
\end{equation}%
is directly related to the symmetry potential $U_{sym}$. Because of the
chemical equilibrium conditions given in Eq. (\ref{coexistencemu}), the
chemical potentials cancel out in the double neutron/proton ratio
\begin{equation}
\frac{(n/p)_{G}}{(n/p)_{L}}(p)=\exp [-2(\delta _{G}\cdot U_{sym}(p,\rho
_{G})-\delta _{L}\cdot U_{sym}(p,\rho _{L}))/T].
\end{equation}%
This general expression clearly demonstrates that the differential IsoF for
energetic nucleons carries direct information about the momentum dependence
of the symmetry potential. In the above expressions, the tiny temperature
dependence of the symmetry potential has been neglected.

For both the liquid-gas and hadron-QGP (quark-gluon-plasma) phase
transitions, equilibrium model calculations for infinite nuclear
matter are very useful for developing new concepts and predicting
novel phenomena. However, the experimental search/confirmation for
the new phenomea/concepts in real nuclear reactions is usually very
challenging. For example, the intermediate energy heavy-ion reaction
community has been studying for more than two decades the underlying
nature and experimental signatures of the LG phase transition
predicted first for infinite nuclear matter based on thermodynamical
considerations. Hopefully, similar to the very fruitful study
recently in momentum-space about how electrons behave during the
phase transition to high-temperature superconductors\cite{Nor04},
the new study about how nucleons behave in the correlated momentum-
and isospin-space may reveal deeper insights into the nature of the
LG phase transition.

\begin{figure}[tbh]
\includegraphics[scale=0.65]{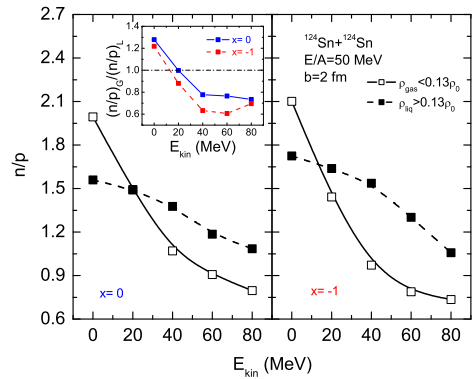}
\caption{{\protect\small (Color online) The neutron/proton ratio in
the \textquotedblleft gas" and \textquotedblleft liquid" phases as a
function of the nucleon kinetic energy for the reaction of
$^{124}$Sn$+^{124}$Sn at $E_{beam}/A=50$ MeV.}} \label{fig3}
\end{figure}
\begin{figure}[tbh]
\includegraphics[scale=0.65]{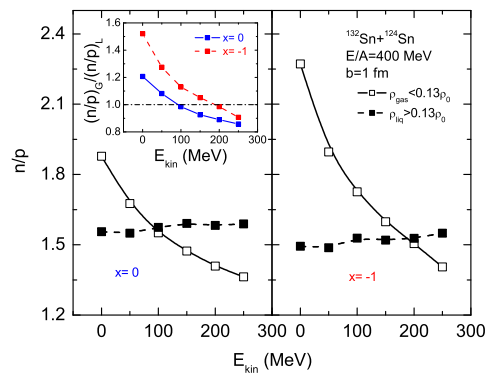}
\caption{{\protect\small (Color online) Same as Fig. 3 but for the
reaction of $^{132}$Sn$+^{124}$Sn at $E_{beam}/A=400$ MeV.}}
\label{fig4}
\end{figure}
How can one experimentally measure the differential IsoF? As for the
structure functions of quarks and gluons in the initial state of
relativistic heavy-ion collisions, the momentum distribution of the
n/p ratio in the liquid phase may not be measured directly since
what can be detected at the end of heavy-ion reactions are free
nucleons and bound nuclei in their ground states. Nevertheless,
precursors and/or residues of the transition in the differential
IsoF may still be detectable in heavy-ion reactions especially those
induced by radioactive beams. While a comprehensive study on the
detailed signatures of the differential IsoF is beyond the scope of
this work, we have actually performed calculations using the IBUU04
transport model with the same MDI interaction\cite{das03}. Shown in
Fig.~\ref{fig3} and Fig.~\ref{fig4} are two typical examples for the
central reaction of $^{124}$Sn$+^{124}$Sn at $E_{beam}/A=50$ MeV and
$^{132}$Sn$+^{124}$Sn at $E_{beam}/A=400$ MeV, respectively. To
separate approximately nucleons in the low density \textquotedblleft
gas" region from those in the \textquotedblleft liquid" region a
density cut at $0.13\rho _{0}$ is used. Calculations using a cut at
$0.5\rho _{0}$ in the reaction at 50 MeV/A lead to qualitatively the
same although quantitatively slightly different results. Very
interestingly, in both reactions there is indeed a transition from
the neutron-richer (poorer) \textquotedblleft gas (liquid)" phase
normally known as the IsoF for low energy nucleons to the opposite
behavior (i.e., anti-IsoF) for more energetic ones. Moreover, the
transition nucleon energy from the normal IsoF to the anti-IsoF is
sensitive to the parameter $x$ used. This is more pronounced in the
reaction at $E_{beam}/A=400$ MeV where effects of the symmetry
(Coulomb) potential are relatively stronger (weaker) for more
energetic nucleons consistent with predictions of the thermal model
in the right window of Fig.\ 2. Comparing the thermal model
predictions and the transport model results, one sees that the two
independent approaches predict qualitatively the same phenomenon
while there are quantitative differences, especially for low energy
nucleons. This is mainly because in nuclear reactions the Coulomb
repulsion shifts protons in the gas phase from low to higher
energies leading to the peak in $(n/p)_G$ ratio at $E_{kin}=0$,
while it has little effects on the protons in the liquid phase.
Transport model calculations without the Coulomb potential lead to
results more closely resembling the thermal model predictions for
infinite nuclear matter. We notice that the ``gas" phase defined in
this study contains also the pre-equilibrium nucleons which are
known to be more neutron-rich than the reaction system\cite{liko97}.
They are energetic and thus affect mostly the high energy part of
the $(n/p)_{G}$ ratio. The subtraction of the pre-equilibrium
nucleons from our analyses will thus mainly lower the $(n/p)_{G}$
for high energy nucleons. Therefore, the transition from the normal
IsoF to the anti-IsoF will become more obvious and it will make our
conclusions even stronger. We also notice that the radial flow
affects all nucleons in both the gas and liquid phases the same way
as already illustrated in ref.\cite{liy}. The subtraction of the
radial flow is expected to have very little effect on the
neutron/proton ratio.

In summary, the differential IsoF is studied as a function of
nucleon momentum. While the gas phase is overall more neutron-rich
than the liquid phase, the gas phase is richer (poorer) only in low
(high) energy neutrons than the liquid phase. Clear indications of
the differential IsoF consistent with the thermodynamic model
predictions are demonstrated within a transport model for heavy-ion
reactions. While it may be very challenging to test experimentally
the predictions, future comparisons with experimental data will
allow us to extract critical information about the momentum
dependence of the isovector strong interaction.

We would like to thank Drs. Wei-Zhou Jiang, Yang Sun and Wolfgang
Trautmann for helpful discussions. This work was supported in part
by the US National Science Foundation under Grant No. PHY-0652548
and the Research Corporation under Award No. 7123, the National
Natural Science Foundation of China under Grant Nos. 10334020,
10575071, and 10675082, MOE of China under project NCET-05-0392,
Shanghai Rising-Star Program under Grant No. 06QA14024, the SRF
for ROCS, SEM of China, and the China Major State Basic Research
Development Program under Contract No. 2007CB815004.

\end{document}